\documentclass[superscriptaddress,twocolumn,aps,pre]{revtex4}
\usepackage{graphicx}% Include figure files
\begin{document}
\title{{\bf Evolutionary Dynamics in Complex Networks of Competing Boolean Agents}}
\author{Baosheng Yuan}
\affiliation{Department of Computational Science, Faculty of Science,
National University of Singapore, Singapore 117543}
\author{Bing-Hong Wang}
\affiliation{Department of Computational Science, Faculty of Science,
National University of Singapore, Singapore 117543}
\affiliation{Department of Modern Physics, University of Science and
 Technology of China, Hefei, Anhui, 230026, China}
\author{Kan Chen}
\affiliation{Department of Computational Science,
Faculty of Science, National University of Singapore, Singapore 117543}

\date{\today }

\begin{abstract}

We investigate the dynamics of network minority games on Kauffman's NK networks (Kauffman nets), growing directed networks (GDNets), as well as growing directed networks with a small fraction of link reversals (GDRNets). We show that the 
dynamics and the associated phase structure of the game depend crucially on the structure of the underlying network. The dynamics on GDNets is very stable for all values of the connection number $K$, in contrast to the dynamics on Kauffman's NK networks, which becomes chaotic when $K>K_c=2$. The dynamics of GDRNets, on the other hand, is near critical. Under a simple evolutionary scheme, the network system with a ``near" critical dynamics evolves to a high level of global coordination among its agents. In particular, the performance of the system  is close to the optimum for the GDRNets; this suggests that criticality leads to the best performance. For Kauffman nets with $K>3$, the evolutionary scheme has no effect on the dynamics (it remains chaotic) and the performance of the MG resembles that of a random choice game (RCG).
\end{abstract}
\maketitle

{PACS numbers: 89.75-k, 89.75.Fb, 87.23.Kg, 05.45.-a}

\section{Introduction}
Complex networks have attracted immense interest in recent years, due to their great capability and flexibility in describing a wide range of natural and social systems. The study of the organization of complex networks has attracted intensive research interest \cite {Albert, Newman, Dorogovtsev} ever since the seminal works of Strogatz on small-world networks \cite{Strogatz} and Barab\'asi and Albert \cite {Barabasi} on scale-free networks. The degree distribution $P(k)$, which is defined as the probability of finding a node with exactly $k$ links, has been considered as the most important characterization of complex networks. The power law or scale-free degree distributions have been established in many real-world complex networks \cite {Strogatz,Barabasi}.

The dynamics of a complex network can be studied in the context of a system of interactive elements (agents) on the network; it depends on how the network is organized 
and how the elements interact. In this paper we study a network version of the minority game (MG) model proposed by Challet and Zhang \cite{Challet}, which is a simplification of Arthur's El Farol bar attendance model \cite {Arthur}. The MG model serves as an interesting paradigm for a system of adaptive agents competing for limited resources. The phase structures of the original MG \cite{Savit} and the evolutionary version of the game \cite {Johnson, Hod, Chen, Yuan1} have been well understood. Note that in the MG models, the agents are not directly linked to one another, but they are influenced by the global environment created by the collective action of all the agents. 

The study of network dynamics is pioneered by Kauffman \cite{Kauffman, Kauffman2} who introduced  NK random networks and studied its Boolean dynamics. Recently there are quite a number of studies on different aspects of network dynamics. Aldana and Cluzel demonstrated that the scale-free network favors robust dynamics\cite{Aldana}. Paczuski \textit{et al} \cite {Paczuski} considered the MG model on a random network to study the self-organized process which leads to a stationary but intermittent state. Galstyan \cite{Galstyan} studied a network MG, focusing on how the change of the mean connectivity $K$ of a random network affects the global coordination of the system of different capacities. 
Anghel \textit{et al} \cite{Anghel} used the MG model to investigate how interagent's communications across a network lead to a formation of an influence network. In this paper we try to address the question of how different network organizations affect the dynamics of the system, which has not been fully explored in the previous studies. We investigate the dynamics of networks in the context of a network minority game. In particular we would like to know 1) how the dynamics and phase structure of the network minority game depend on the network organization, and 2) how evolution affects the dynamics and phase structure of the game. We will consider three types of networks: Kauffman's NK random networks (Kauffman nets), growing directed networks (GDNnets), and growing directed networks with a fraction of link reversals (GDRNets) as described in Ref.~\cite{Yuan2}. 

The paper is organized as follows. In the next section, we will describe the network MG model and the various networks we consider in this paper. Section III presents our numerical results and some qualitative analysis. The last section is the summary.

\section{Network Minority Game Model}
\subsection{MG of network agents}
The network based MG model consists of $N$ (odd number) agents described by the state variables $s_i=\{0,1\}, i=1,2,...,N$, each connected to another $K$ agents, $i_1, i_2,...,i_K$.
Each agent has $S$ strategies which are the mapping functions specifying a binary output state (0 or 1) for each possible input vector consiting of the states of her $K$ connected agents.  
The state or decision of the $i$th agent at the current time step $t$ is determined by the states/decisions of the $K$ agents it connects to at the previous time step $t-1$, i.e.
\begin{equation} \label{eq:strategy} 
s_i(t)=F_i^j(s_{i_1}(t-1), s_{i_2}(t-1),...,s_{i_K}(t-1))
\end{equation}
where $s_{i_k} (k=1,2,...,K$) is the state of the $k$th agent that is connected to agent $i$, and $F_i^j, j=1,..,S$ are $S$ Boolean functions (strategies) taken from the strategy space consisting of $2^{2^K}$ strategies.
As in the standard minority game, each agent keeps a record of the cumulative wealth $W_i(t)$ as well as the cumulative (pseudo) scores, $Q_i^s(t), i=1,2,...,N, s=1,2,..,S$, for each of her $S$ strategies.
Before the game starts each agent selects at random one of her 
$S$ strategies, and the cumulative wealth $W_i(t)$ and strategy scores $Q_i^s(t)$ are initialized to zero. At each time step, each agent decides which of the two groups (0 or 1) to join based on the best-scoring strategy (the strategy that would have made the most winning predictions in the past) among her $S$ strategies.
Each agent gains (loses) one point in her cumulative wealth for her winning (losing) decision and each strategy gains (loses) one pseudo-point for its winning (losing) prediction. The agents who are among the minority win; those among the majority lose. 
Let $A(t)$ be the number of agents choosing $1$ at time step $t$. Then a measure of utilization of the limited resources (system performance) can be defined as the variance of $A(t)$ over a time period $T$:
\begin{equation} \label{eq:variance} 
\sigma ^2 = \frac{1}{T} \sum_{t=t_0}^{t_0+T} (A(t) - \bar{A})^2
\end{equation}
where $\bar{A}=\frac{1}{T} \sum_{t=t_0}^{t_0+T} A(t) \sim N / 2$ is the mean number of agents choosing $1$.
Clearly $\sigma ^2$ measures the global coordination among agents. The optimal (smallest) value of variance $\sigma ^2 $ is $0.25$, where the number of winning agents reaches its optimal value, $(N-1)/2$, in every time step. 
For a random choice game (RCG), where each agent makes decision by coin-tossing, the value of the variance $\sigma ^2$ is $0.25 N$. The game is adaptive as each agent has $S$ strategies to choose from, attempting to increase her chance of winning.  

It's worthwhile to point out the differences between the network MG we discuss here and the original MG defined in Ref.~\cite{Challet}. In the original MG, the input for each agent's strategies at time $t$ is a vector of the winning decisions of the game in the previous $M$ time steps.
In the network (local) MG, however, the input for each agent's strategies at time $t$ is a vector consisting of the decisions of the $K$ agents she connects to at the previous time step $t-1$.
Thus, the first key difference is that the agents in the original MG use global information while the agents in the network MG use local information. The second key difference is that the original MG is based on the $M$ time-step history of global information while the network MG employs a one-step forward dynamics.

\subsection{Evolutionary Minority Game}
We first investigate the adaptive MG models and used the results as a basis for comparison with the evolutionary MG model, which is our main interest. In an evolutionary dynamics, the quenched strategies can be changed and more generic behaviors will emerge. We use the same evolutionary dynamics as the one described in Refs.~\cite{Johnson,Yuan1}. In this scheme, each agent is required to change her $S$ strategies (by choosing $S$ new strategies randomly) whenever her cumulative wealth $W_i(t)$ is below a pre-specified bankruptcy threshold, $-W_c (W_c > 0)$.  
The bankrupted agents re-set their wealth and strategy pseudo-scores to zero, and the game continues.  The network connection, however, does not evolve. We have found that this evolution scheme is much more effective than the scheme used,
for example, in Ref.~\cite {Paczuski} where the evolution happens at the end of every epoch of specified duration (say $10,000$ time steps), and only the worst performer is required to change her strategies after each epoch.

\subsection{Generation of Different Networks}
The dynamics of the game depends crucially on the network organization. In our model, the network connection, once generated, remains unchanged throughout the game.\\

\noindent\textbf{The Kauffman NK random network (Kauffman net)}\\
There are a few different ways to generate random networks \cite{Erdos, Watts}. To study dynamics, however, it is
more convenient to use Kauffman's NK random networks. The Kauffman net is generated by specifying $N$ agents first, and then connecting each agent randomly to $K$ other agents, whose decisions serve as the input to its strategies. \\

\noindent\textbf{\textbf{Growing directed network (GDNet)}}\\ 
A growing directed network can be generated as described in Ref.~\cite{Yuan2}. We start with an initial cluster of $K+1$ agents, which are mutually connected (two directed links between each pair of agents). At each stage, we add a new agent and connect it to $K$ other agents already present in the network. The link is directed from the new agent to the existing ones, meaning that the strategies of the new (younger) agent are based on the states of the existing (older) ones.
We assume that the probability of connecting a new agent to an existing one with degree $k_{in}$ is proportional to $k_{in}^{\alpha}+1$, where $k_{in}$ is the number of incoming links to the existing agent. The constant 1 is added to give a nonzero starting weight to the agents that have not been connected to.
For $\alpha = 0$, we have a \textit{growing directed random network} which we refer to as \textbf{GDNet I}.
For $\alpha > 0$, we have preferential attachment. The special case of $\alpha = 1$ corresponds to a \textit{scale-free directed network} which we refer to as \textbf{GDNet II}.
In this network, the out-degree is $K$ for all the agents, but the in-degree follows a power law distribution.  The undirected version of this model corresponds to the Barab\'asi-Albert model.
In the growing networks (random or scale-free), the younger agents are influenced by the older ones, except for the initial $K+1$ agents who are mutually influenced.\\

\noindent\textbf{\textbf{Growing directed network with a fraction of link reversals (GDRNet)}}\\
 In order to make our discussion more relevant to many real world networks which typically have a fraction of feedback links, we here modify the above growing network to allow a small fraction of link reversals. 
Let $p$ be the probability that each agent has a link reversal: when each new agent is connected to other  $K$ agents  already present in the network, each link has a probability of $q=p/K$ to have its direction reversed.
We consider two GDRNets: GDRNet I with $\alpha = 0$ and GDRNet II with $\alpha = 1$.

There are two new features for GDRNets that are worth mentioning: 1) Some agents may have more than $K$ strategy inputs, while others may have fewer than $K$ inputs; but the mean number of inputs for an agent remains as $K$. 
2) Some younger agents can influence the older agents.

\section{Numerical Results and Analysis}

\subsection{MG on the Kauffman net}

Let's first examine the performance of MG on the Kauffman net. The result is shown in Fig.~1. It is clear from the figure that when $K=2$, the variance $\sigma ^2$ has very large fluctuations (four orders of magnitude for $N=401$ and five orders of magnitude for $N=901$) with different initial conditions.  
This reflects the fact that the system dynamics is critical  for $K=2$.
For $K \ge 3$ the system performs like a random choice game.
The observation we obtained here is consistent with the well-known result for the Boolean dynamics on Kauffman nets:
when $K=2$ the system is at the ``edge of chaos'' and for $K \ge 3$ the system is chaotic~\cite{Kauffman,Derrida}.

\begin{figure}
\begin{center} \includegraphics[height=4.6cm, width=8.5cm]{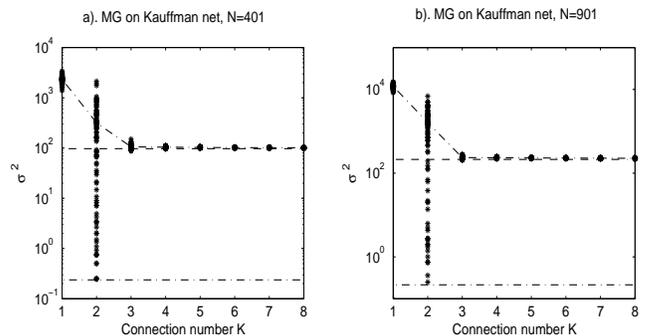} \end{center}
\caption {The variance $\sigma ^2$ of the network MG as a function of $K$ on \textbf{Kauffman NK random network}. 
The dash-line in the middle is for $\sigma^2=0.25N$ (RCG), and the dash-dot line at bottom is for  $\sigma^2=0.25$ (the theoretical lower bound). 
The system parameters are: $S=2$, $N=401$ (left plot) and $901$ (right plot). 
The figures are obtained with $100$ simulations each.}
\end{figure}

The dynamics in the original MG depends on two variables: $N$, the system size and $M$, the memory size of the agents.
There are three different phases for different memory value $M$, described by a Savit curve~\cite{Savit}.
The critical value for an optimal global coordination is $M_c \sim \ln(N)$, which depends on $N$. For the network MG on the Kauffman net, however, the three phases of the dynamics are: stable for $K=1$, critical  for $K=2$, and chaotic for $K \ge 3$.
So the dynamics of network MG depends on only one variable, $K$, and the chaotic regime dominates.

\subsection{EMG on the Kauffman net}

\begin{figure}
\begin{center} \includegraphics[height=4.6cm, width=8.5cm]{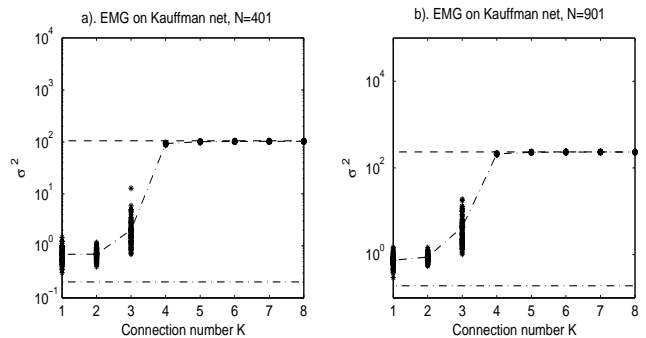} \end{center}
\caption{The variance $\sigma ^2$ of the network EMG on the Kauffman net as a function of $K$.
The dash-line in the middle is for $\sigma^2=0.25N$ (RCG), and the dash-dot line at bottom is for  $\sigma^2=0.25$.
The system parameters are: $S=2$, $N=401$ (left plot) and $901$ (right plot), and $W_c =1024$.
The figures are obtained with $100$ simulations each.}
\end{figure}

Now let's check how the game performs when the simple evolution scheme described above is applied. 
The results are shown in Fig.~2.
We see from the figure that evolution helps dramatically improve the system performance when the connection number $K$ is samll ($K \le 3$), but has virtually no effect for larger $K(>4)$.  
This means that for $K \le 3$ the system is still at stable or ``critical'' state, but for $K \ge 4$ it is  chaotic. Note that evolution has shifted the critical point from $K=2$ to $K=3$, suggesting that
it is more powerful than adaptation (modeled by strategy switching) in bring out order in complex systems. 
Since most real-world systems evolve, evolutionary dynamical analysis may be more relevant.

\subsection{MG on Growing Directed Networks}

Let's now examine the performance of MG on growing directed networks.  Two limiting cases of the growing directed networks are checked: 1) \textbf{GDNet I}, the growing random directed network ($\alpha = 0$);  2) \textbf{GDNet II}, the growing directed network with preferential attachment ($\alpha=1$). Fig.~3 shows the results of system performance.

\begin{figure}
\begin{center} \includegraphics[height=8.5cm, width=8.5cm]{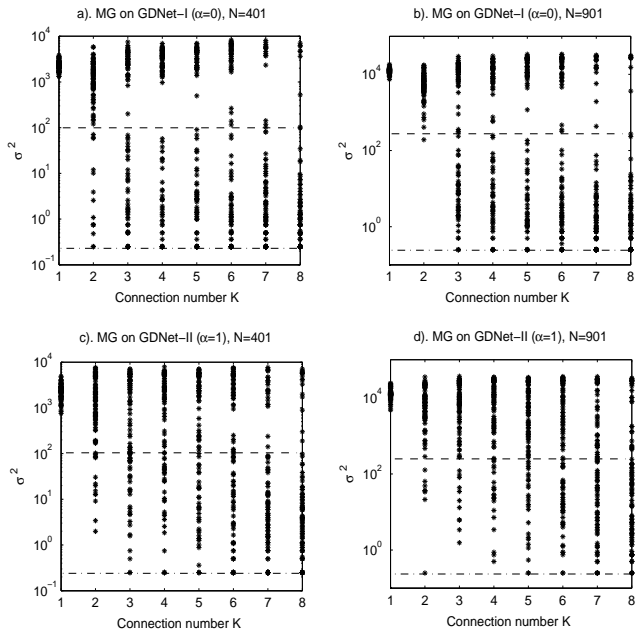} \end{center}
\caption{The variance $\sigma ^2$ of the network MG as a function of $K$ on \textbf{GDNet I} (upper panel) and \textbf{GDNet II} (bottom panel). 
The dash-line in the middle is for $\sigma^2=0.25N$ (RCG), and the dash-dot line at bottom is for  $\sigma^2=0.25$.
The system parameters are: $S=2$, $N=401$ (left plot) and $901$ (right plot). The figures are obtained with $100$ simulations each.}
\end{figure}

We have the following observations for the MG dynamics: 1) there are large fluctuations in the values of the variance $\sigma^2$;  2) there seems to be no significant difference in the MG dynamics for the two growing network models.
So in terms of a simple (non-evolutionary) dynamical process, all growing networks (irrespective of its value of preferential attachment exponent $\alpha \in [0,1]$) have similar dynamics and the scale-free network is not special.
The stability of the dynamics is due to the construction process of growing networks, which leads to a maximum state cycle length of $2^{K+1}$ as was pointed out in Ref.~\cite{Yuan2}, irrespective of the value of $\alpha$. 

Comparing the dynamics of growing networks with that of the Kauffman net, we see a lot of differences. In the Kauffman net (Fig.~1) the dynamics is  stable or critical for $K \le 2$. In growing networks (Fig.~3), however, the dynamics is stable for all the values of $K$, on both \textbf{GDNet I} and \textbf{GDNet II}.
Thus non-growing and growing networks have very different network dynamics.

\subsection{EMG on Growing Directed Networks}

\begin{figure}
\begin{center} \includegraphics[height=8.5cm, width=8.5cm]{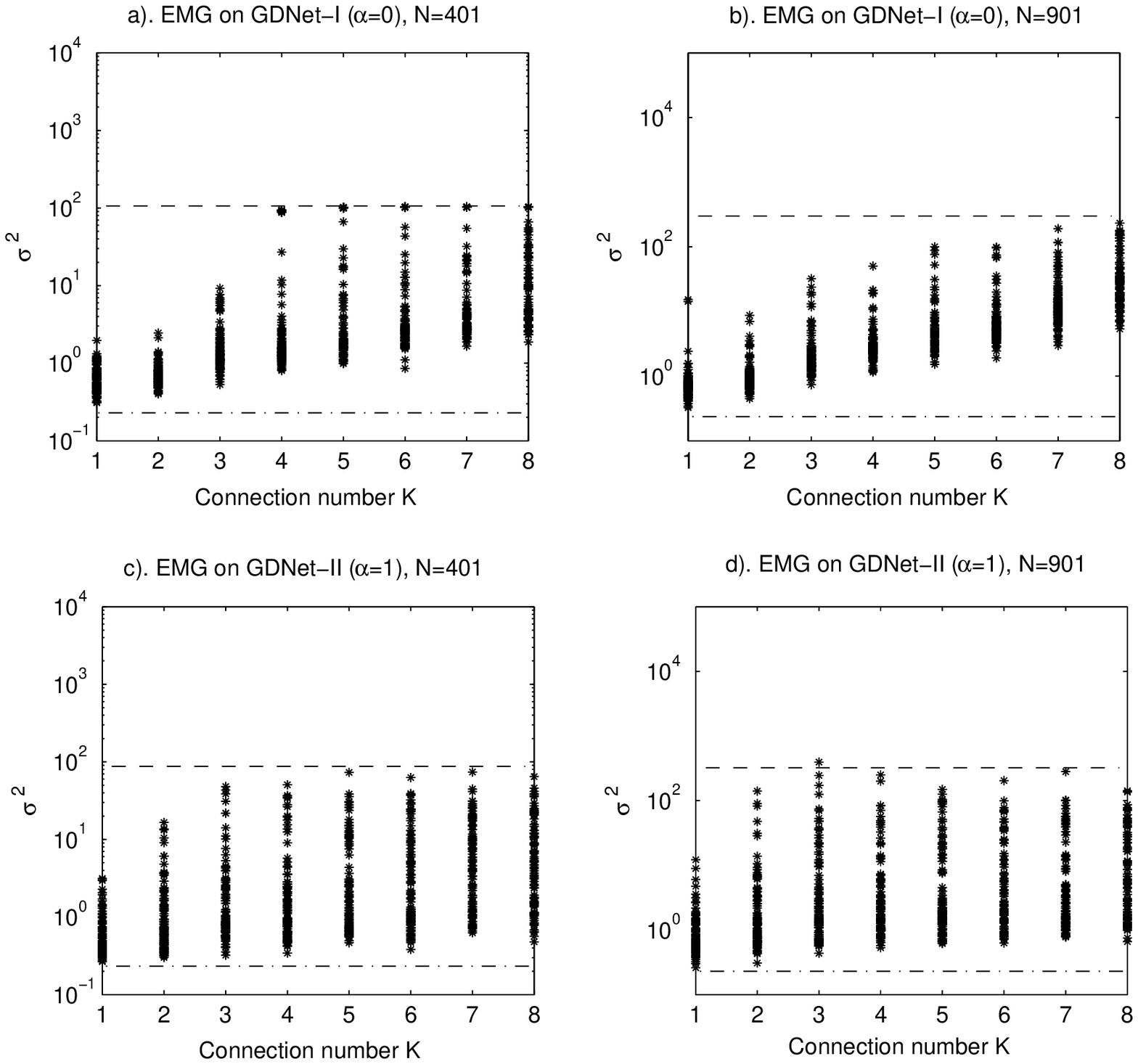} \end{center}
\caption{The variance $\sigma ^2$ of the network EMG as a function of $K$ on \textbf{GDNet I} (upper panel) and \textbf{GDNet II} (bottom panel). 
The dash-line in the middle is for $\sigma^2=0.25N$ (RCG), and the dash-dot line at bottom is for  $\sigma^2=0.25$.
The system parameters are: $S=2$, $N=401$ (left plot) and $901$ (right plot), and $W_c =1024$. The figures are obtained with $100$ simulations each.}
\end{figure}
To see how evolution affects the dynamics, we plot the results on system performance for the EMG on GDNets in Fig.~4.
By comparing Fig.~3 and Fig.~4, we see that evolution helps reduce the variance $\sigma^2$ dramatically (by more than two orders of magnitude). 
Although the fluctuations in the variance is still large, the values are all below the value of the variance corresponding to RCG. We can also see (by comparing panels a-b to panels c-d in the Fig.4) that the EMG on GDRNet I performs better than the EMG on GDRNet II, but the difference is small.
This is not surprising as the essential property of all growing networks are the same: the younger agents are always influenced by the older ones; and the system state cycle is determined only by the initial cluster of $K+1$ agents, leading to a stable dynamics of the system with the state cycle length limited to the maximum length of $2^{K+1}$.

By comparing the results for growing networks (shown in Figs.~3-4) with the results for the Kauffman net (shown in Figs.~1-2), we see that these two types of networks have very different impacts on the dynamics. 
The reason for this is rooted in the differences in network construction process.
For the Kauffman net, each agent chooses, at random, other $K$ agents for inputs to her strategies. Any given agent has a potential to influence many other agents. It is not surprising that, for large enough $K$ ($K \ge 3$), the system  is virtually chaotic. However, for GDNets (with any value of $\alpha \in [0,1]$), the dynamics is driven by the initial cluster of $K+1$ agents; this results in a stable dynamics in which the maximum cycle of length is $2^{K+1}$.

\subsection{MG/EMG on Growing Networks with Link Reversals}

Now we examine the performance of the MG on growing networks with link reversals, \textbf{GDRNet I} and \textbf{GDRNet II}.  The results for the MG are shown in Fig.~5 and the results for the EMG are shown in Fig.~6.

\begin{figure}
\begin{center} \includegraphics[height=8.5cm, width=8.5cm]{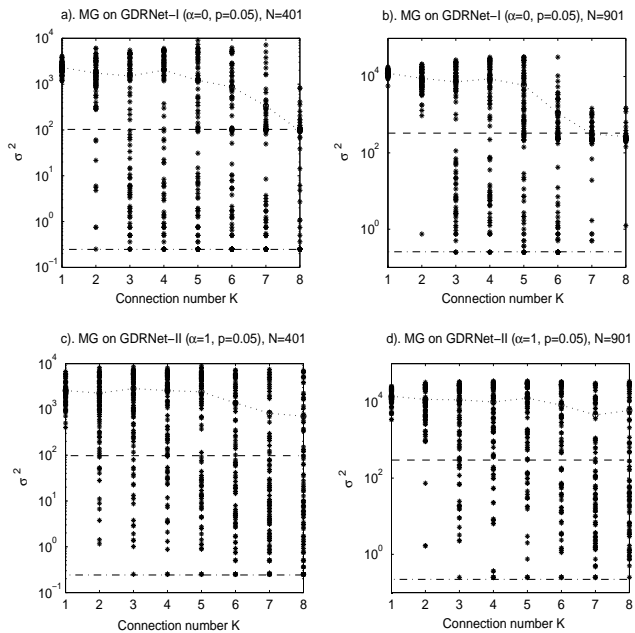} \end{center}
\caption{The variance $\sigma ^2$ of the MG as a function of $K$ on \textbf{GDRNet I}(upper panel) and \textbf{GDRNet II} (bottom panel). 
The link reversal probability is $q=p/K$ (here $p=0.05$ is the probability that an agent has a reversed link). 
The dot-line at the top is the average $\bar{\sigma^2}$, the dash-line in the middle is for $\sigma^2=0.25N$ (RCG), and the dash-dot line at bottom is for $\sigma^2 =0.25$. 
The other system parameters are: $S=2$, $N=401$ (left plot) and $901$ (right plot). The figures are obtained with $100$ simulations each.}
\end{figure}

\begin{figure}
\begin{center} \includegraphics[height=8.5cm, width=8.5cm]{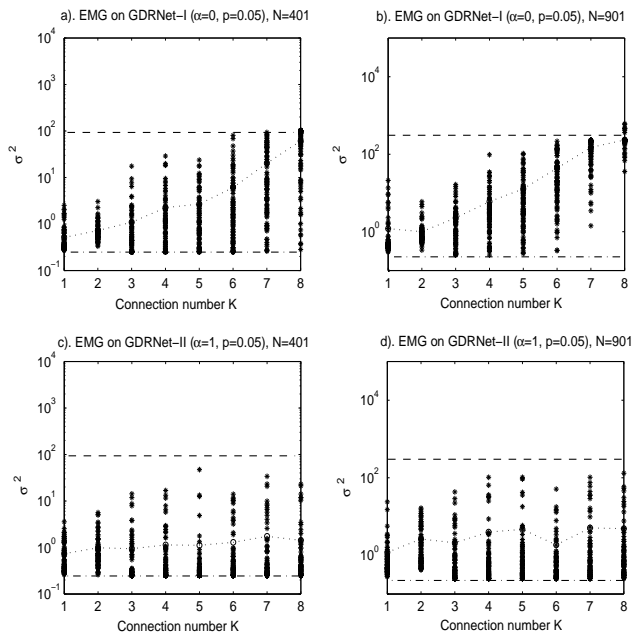} \end{center}
\caption{The variance $\sigma ^2$ of the EMG as a function of $K$ on \textbf{GDRNet I} (upper panel) and \textbf{GDRNet II} (bottom panel). The link reversal probability is $q=p/K$ ($p=0.05$). 
The dash-line at the top is for $\sigma^2=0.25N$ (RCG), the dot-line below is the average $\bar{\sigma^2}$, and the dash-dot line at bottom is for $\sigma^2 =0.25$. 
The other system parameters are: $W_c =1024$, $S=2$, $N=401$ (left plot) and $901$ (right plot). 
The figures are obtained with $100$ simulations each.}
\end{figure}

Fig.~5 shows that, without evolution, the MG on \textbf{GDRNets} produces similar results as the MG on \textbf{GDNets}. However, the EMG on \textbf{GDRNets} produces significantly better results than the EMG on \textbf{GDNets}, as shown in Fig.~6. \textbf{GDRNet II}, in particular, gives the best performance among all the network models. 
These observations suggest that MG dynamics is not very sensitive to different attachment algorithm in a growing directed network. But the attachment algorithm makes some difference in the performace of the EMG.
If we examine the results (shown in Fig.6) more carefully we see that for \textbf{GDRNet II}, the EMG performance is so good that the variance $\sigma^2$ is below $1$  most of the time.
This means that the difference between the attendance numbers in the majority group and the minority group is less than $2$ on average, which is very close to the theoretical bound where the difference is $1$ at every time step.

We have shown in our earlier paper \cite{Yuan2} that, the general Boolean dynamics on GDRNets is close to  critical, in the sense that the distribution of the state cycle lengths is close to a power law. This suggests that, under an evolutionary dynamics, criticality makes the system more efficient.

\section{Summary}
We have presented an extensive numerical investigation on the dynamics of the MG on various networks.
We have compared the dynamics of the network MG with that of the original MG. In the original MG, the critical value for an optimal global coordination is $M_c \sim \ln(N)$, so the dynamics depends on two variables: $N$, the system size and $M$, the memory size of the agents.
In the network MG, however, the dynamics depends on $K$ only, and the system has three phases: stable for $K=1$, critical for $K=2$, and chaotic when $K \ge 3$, consistent with Kauffman's Boolean dynamics.
We have studied the dynamics of the EMG on the Kauffman net, and we have found that, it appears to be ``critical''  for the connection number $K = 3$, which is different from the critical value $K_c = 2$ for the non-evolutionary MG. This shows that evolution makes a significant difference.

We have investigated the dynamics of the MG on a general class of growing directed networks.  We show that, the dynamics on these networks is stable on these growing networks, which is very different from the dynamics on the Kauffman NK random network. This is due to the way the network is constructed. 
In the Kauffman net all the agents are treated equally; every agent has an equal probability to influence others. 
This results in a very large ``influence network" of a given agent. 
However, in growing directed networks, the initial cluster of agents dictate the dynamics of the system; the ``junior'' agents have no influence on the ``senior'' ones.  The maximum  state cycle length is limited to  $2^{K+1}$.

We have also studied the MG dynamics for a modified growing directed network model which allows a small fraction of link reversals.
Our numerical results show that the best system coordination and performance emerges for the EMG on the scale-free network with link reversals (\textbf{GDRNet II});  the variance $\sigma^2$ in the EMG on \textbf{GDRNet II} reaches to such a low level that it's close to the theoretical bound most of the time. This suggests that evolution makes the agents best coordinated on critical scale-free networks.

\end{document}